# Microscale adhesion patterns for the precise localization of amoeba

Tzvetelina Tzvetkova-Chevolleau [a1], Edward Yoxall [a,b], David Fuard [a], Franz Bruckert [b], Patrick Schiavone [a], Marianne Weidenhaupt [b]

*a Laboratoire des Technologies de la Microélectronique CNRS, 17 rue des Martyrs, 38054 Grenoble, France*

*b Laboratoire des Matériaux et du Génie Physique, INPGrenoble Minatec - 3 parvis Louis Néel – 38016 Grenoble, France*

**Abstract**

In order to get a better understanding of amoeba-substrate interactions in the processes of cellular adhesion and directional movement, we engineered glass surfaces with defined local adhesion characteristics at a micrometric scale. Amoeba (*Dictyostelium dicoideum*) are capable to adhere to various surfaces independently on the presence of extracellular matrix proteins. This paper describes the strategy used to create selective adhesion patterns using an appropriate surface chemistry and shows the first results of locally confined amoeba adhesion. The approach is based on the natural ability of *Dictyostelium* to adhere to various types of surfaces (hydrophilic and hydrophobic) and on its inability to spread on inert surfaces, such as the block copolymer of polyethylene glycol and polypropylene oxide, named Pluronic. We screened diverse alkylsilanes, such as methoxy, chloro and fluoro silanes for their capacity to anchor Pluronic F127 efficiently on a glass surface. Our results demonstrate that hexylmethyldichlorosilane (HMDCS) was the most appropriate silane for the deposition of Pluronic F127. A complex dependence between the physico-chemistry of the silanes and the polyethylene glycol block copolymer attachment was observed. Using this method, we succeed in scaling down the micro-fabrication of pluronic-based adhesion patterns to the amoeba cell size (10µm). This original pluronic patterning method should prove useful as a tool for controlling cell adhesion and directional movement in amoeba.© 2008 Elsevier Science. All rights reserved

## 1. Introduction

Cell-substrate dependant polarization and directional movement are crucial for many physiological processes such as embryonic development, wound healing and functional immune [1] and neural [2] systems.

The ability to engineer adequate chemical surfaces for specific cell-surface interactions like spatial control over cell polarization and directional movement has tremendous potential. It opens the way to controlled and predictable host biomaterial interactions that can find an application in tissue engineering and in the development of new biomedical implant materials.

cell adhesion and movement (for a review, see [4]). In order to control surface adhesion spatially we

In order to study polarization and cell movement on specific adhesive patterns, we selected the commonly used eukaryotic model cell *Dictyostelium discoideum*, a social amoeba. It is amenable to routine genetic techniques (directed and random mutagenesis, deletion and over-expression of genes, complementation) and its genome has been fully sequenced [3]. Moreover many of the *Dictyostelium* genes show a high degree of sequence similarity to genes in vertebrate species and the molecular machineries driving adhesion and cell movement are broadly conserved. These characteristics together with the ease of its culture conditions make *Dictyostelium* an organism of choice for the study of made use of *Dictyostelium's* natural capacity to adhere to almost all types of surfaces and its inability to adhere to special, so-called inert surfaces, such as the block copolymer of polyethylene glycol and polypropylene oxide, named Pluronic [5].

---

[1] Corresponding author. Tel.: +33-43-878-4812; fax: +33-43-878-9292; e-mail: tzvetelina.tzvetkova@cea.fr

The paper describes the surface chemistry used to create selective adhesion patterns. Different surface characterization results used to select these surface treatments are discussed.

## 2. Experimental details

*2.1. Pattern microfabrication and surface functionalization.*

Cleaned glass cover slips ($25 \times 60$ mm) were washed with ethanol in an ultrasonic water bath during 5 min. The slides were treated with oxygen plasma (Fig.1a), 100 sccm $O_2$, in a LAM 9400 SE reactor at 50mT for 1 min in order to render the glass very hydrophilic (water contact angle in the range of 10° to 20°). The surfaces were covered with positive UV-sensitive resist, S1813 (Fig.1b) (Shipley Company Inc, USA). The resist was spin-coated and cured according to the manufacturer's protocol to form a uniform UV-sensitive film of 6.5μm thickness. Standard contact photolithography (Fig.1c) using 4-inch chromium masks was then used (UV light: Karl Süss aligner MJB4, SUSS MicroTec, Saint-Jeoire, France at 365 nm and 100 mJ/cm$^2$) and the irradiated pattern was revealed with MF-319 developer (Fig.1d).

An efficient deposition of Pluronic F127 requires surfaces with water contact angles of about 80° on which Pluronic F127 forms an inert monolayer [5, 6]. On the opposite, a hydrophilic surface induces formation of surface aggregates [7] that are weakly attached to the substrate surface. Thus on hydrohpilic surfaces Pluronic F127 is little or not at all adsorbed [6]. Consequently, after the removal of the unpolymerized resist, a hydrophobic layer (Fig.1e) was added by vapor deposition using diverse silanes (as listed bellow) or $CH_4$ or $C_4F_8$ plasma treatment. This resulted in the formation of either hydrophobic surfaces (water contact angles in the range of 90°-110°) or slightly less hydrophobic surfaces (water contact angles in the range of 75°-90°) (Fig.1f), depending on the type of silane used. In the end, the polymerized resist was dissolved with acetone in an ultrasonic water bath during 20 min (Fig.1g), which revealed the hydrophilic pattern under the resist protection (water contact angle in the range of 20° to 30°). The glass slides obtained were treated for 30 min with the anti-adhesive, nonionic copolymer surfactant pluronic F127 (Sigma-Aldrich, 1mg/ml in water) (Fig.1i).

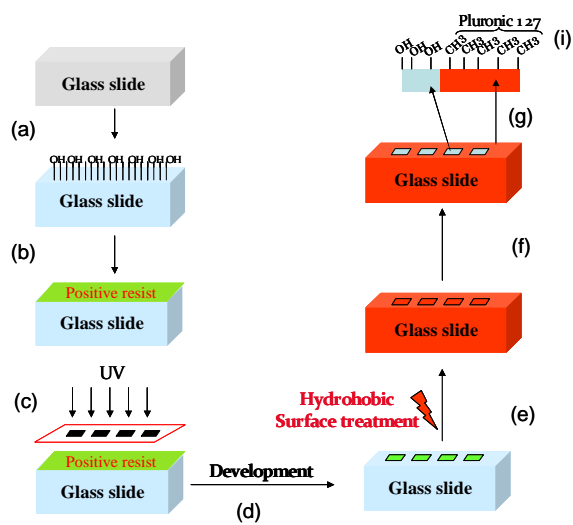

Fig.1 Strategy depicting consequential steps of surface chemistry for the micro-fabrication of adhesive patterns: (a) oxygen plasma treatment of the glass slides; (b) UV-sensitive resist deposition; (c) standard contact photolithography and resist development (d); (e) hydrophobic surface treatment; (f) aquisition of highly hydrophobic glass slide surfaces; (g) acetone removal of the polymerized resist; (i) Pluronic F127 deposition.

*2.2. Chemicals and plasma treatments.*

To obtain highly hydrophobic glass surfaces required for Pluronic F127 deposition, we treated the glass slides in a covered glass chamber for 3 min at 100°C with diverse silanes supplied by ABCR (Karlsruhe, Germany) as follows:
MTMS:Methyltrimethoxysilane;
DMCS:Dimethylchlorsilane;
HMDS :Hexylmethyldichlorsilane;
OTCS :Octadecyltrichlorosilane
PFES : Perfluorodecyltriethoxysilane
PFCS : Perfluorodecyltrichlorooxysilane
TFHS : Tridecafluorotetrahydrooctyldimethylchlorosilane;
2MPEPS: 2-methoxy(propyethoxylenoxy)propyltrimethoxysilane.

Similarly, deposition of a commercial fluoro silane product, OPTOOL DSX$^{TM}$ from Daikin was performed according to the manufacturer's protocol.

Alternatively, we treated the glass slides with $CH_4$ plasma (30 sccm $CH_4$ with 100 sccm Ar, in a LAM 9400 SE reactor at 10mT for 5sec) or $C_4F_8$ plasma (30 sccm $C_4F_8$ with 100 sccm Ar, in a LAM 9400 SE reactor at 10mT for 5 sec). It has been reported that both plasma treatments result in the formation of methyl and teflon-like layers, respectively [8, 9].

*2.2. Drop angle and surface energy measurements*

Contact angles were measured on solid substrates at room temperature using the sessile drop method with the drop shape analysis system G10/DSA10 (Krüss, Germany) using three different liquids: di-iodomethane, ethylene glycol and water. The average values of contact angles were determined from at least 4 droplets of each liquid. Using the extended Fowkes method and taking in consideration the measures of the drops contact angles on the substrate we have derived the three surface energy fractions: disperse, electrostatic and hydrogen bond fraction [10].

*2.3. Cell culture*

*Dictyostelium discoideum* strain AX-2 cells were grown in HL5 medium with agitation (180 rpm) at 21°C. Vegetative cells were harvested by centrifugation (1000 × g, 4°C, 4 min) and resuspended in Sörensen phosphate buffer (pH = 6.2). Standard glass slides containing adhesive patterns were installed in a 10 cm Petri dish and covered with dense cell suspension ($10^7$ cells). Cells were imaged at 2.5x magnification using an Olympus IX-71 inverted microscope and ImagePro imaging software.

3. **Results and discussion**

*3.1. Surface energy measurements*

Fig.2. shows the results of surface energy measurements as a breakdown in the 3 surface energy components for the various treatments used. We observe that the fluoro-silanes display a relatively weaker total surface energy (of about 35mN/m) when compared to the other groups of silanes used in this study, methoxy and chloro silanes (45mN/m and 40mN/m, respectively). This is mainly due to an important reduction in the dispersive energy component. All silane treatments resulted in a reduction of the hydrogen bonding energies when compared to the control, which suggests that these surfaces are less favourable to hydrogen binding reactions. Both types of surface energies, the dispersive and the hydrogen bonding are closely linked to the increase of surface hydrophobicity after silane treatment.

Concerning the electrostatic energies that are indicators of the presence of charges on the surface and so of its capacity to sustain electrostatic interactions, we observed no correlation between the type of silane used and this energy component.

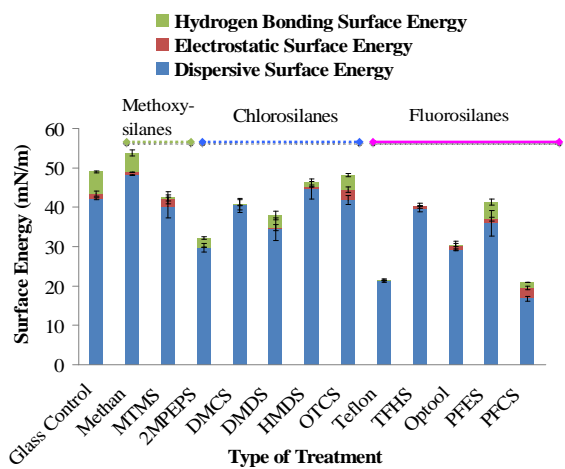

Fig.2 Surface energy breakdown for various surface treatments.

*3.2. Efficiency of the Pluronic F127 deposition*

Polyethylene glycol and its block copolymers are known since long for their cell antiadhesive characteristics [5, 6]. Therefore, we used the quantification of attached cells as an indirect measure of the presence of Pluronic F127. In order to do this, we treated glass slides with different silanes on one half of the slide, leaving the other half hydrophilic. Such slides were then used with and without Pluronic F127 to count the number of cells adhering to each. On Fig. 3 cell adhesion ratios on hydrophilic vs. hydrophobic surfaces are presented with respect to the presence or absence of Pluronic. We observe that HMDCS shows the highest hydrophilic vs. hydrophobic cell adhesion ratio after Pluronic treatment, which suggests an efficient anchoring of the anti adhesive Pluronic F127 and thus selective cell adhesion. Indeed, a limited number of cells

(bellow 15%) is attached to the Pluronic F127 treated HMDCS surfaces. Pluronic attachment is essential for prevention of *Dictyostelium* cell adhesion as illustrated by the cell adhesion ratios on hydrophilic vs hydrophobic surfaces without Pluronic F127 treatment.

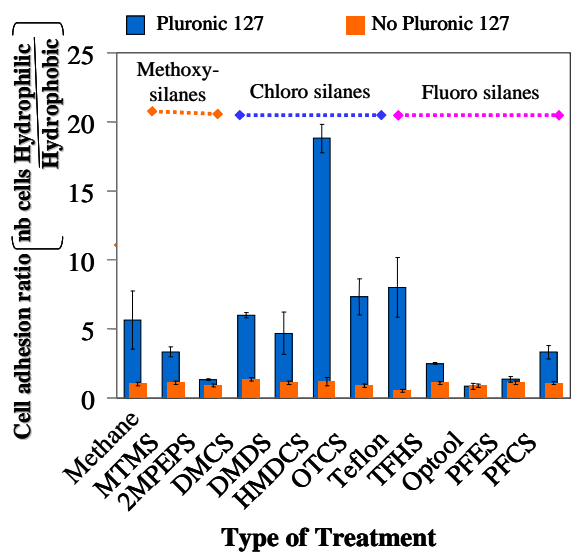

Fig. 3. Efficiency of Pluronic F127 deposition measured as a function of cell adhesion.

Earlier studies [6, 11] have reported that Pluronic deposition is fairly dependent on the substrate hydrophobicity. In this study we observe that an increase in the hydrophobicity of the substrate, which is linked to a decrease in its hydrogen bonding energy component, is a necessary but not a sufficient condition for Pluronic F127 attachment and consequently prevention of *Dictyostelium* adhesion. Indeed Optool, which has a lower hydrogen bonding energy (contact angle of 105°±1°) when compared to HMDCS (contact angle of 97°±2°)), did not result in stable Pluronic F127 attachment as shown by the respective cell adhesion ratios on Fig. 3.

*3.3 Pattern microfabrication*

The micropatterning method depicted on Fig. 1 was used to create Pluronic F127 line patterns on HMDCS treated glass. Amoeba cells were put on these engineered glass surfaces and selective cell attachment on inert zones (containing a hydrophobic HMDCS pattern) was analyzed. The dimensions of the patterned lines were progressively reduced from mm to µm scale. As can be seen on Fig. 4, we succeeded to scale the adhesive (hydrophilic) surfaces down to amoeba cell size (10µm) with good spatial adherence selectivity. Analysis of the cell distribution within the patterned regions revealed that 85% of the cells are adhering to the hydrophilic, adhesive area missing Pluronic, wile only 15% of the cells adhere to the inert Pluronic-treated hydrophobic area. This is in accordance with the results obtained from the efficiency assessment of the deposition of HMDCS and Pluronic on the mm scale (Fig.3).

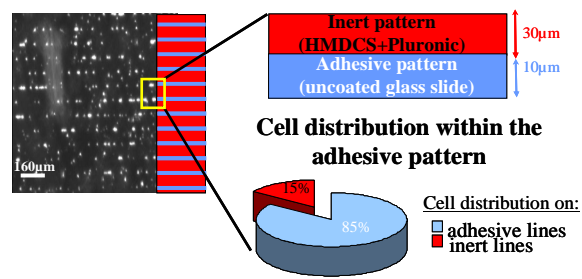

Fig. 4. Selective adhesion of amoeba on linear patterns created using HMDCS and Pluronic F127

## 4. Conclusion and perspectives

In this work we demonstrate that HMDCS is the most appropriate silane in our system for efficient Pluronic F127 attachment. The surface chemistry described here uses HMDCS and Pluronic F127 to selectively prevent amoeba adhesion. We succeed to scale down the micro-fabrication of the adhesive patterns to the amoeba cell size (10µm). In the future we will apply this method to create asymmetric adhesion patterns at the micron scale. Analysis of cell polarization and movement on such patterns will be used to identify distinct molecular driving forces of adhesion-induced cell movement. Polarization inducing adhesion patterns should permit to control the directional movement of cells. Consequently they can be adapted and used to screen and isolate mutants incapable to sense and respond correctly to the asymmetry on the adherent surface.